\documentclass[twocolumn,epjc3]{svjour3}          

\RequirePackage[T1]{fontenc}

\RequirePackage[pdftex]{graphicx}
\RequirePackage{mathptmx}      
\RequirePackage{flushend}
\RequirePackage[numbers,sort&compress]{natbib}
\RequirePackage[colorlinks,citecolor=blue,urlcolor=blue,linkcolor=blue]{hyperref}
\RequirePackage{microtype}
\RequirePackage{amsmath}
\journalname{Eur. Phys. J. C}

\begin{document}

\title{Non-Gaussian Saha's ionization in Rindler spacetime and the equivalence principle}


\author{L. L. Sales\thanksref{e1,addr1,addr2}
        \and
        F. C. Carvalho\thanksref{e2,addr1}  
}

\thankstext{e1}{lazarosales@alu.uern.br}
\thankstext{e2}{fabiocabral@uern.br}

\institute{Departamento de F\'\i sica, Universidade do Estado do Rio Grande do Norte, 59610-210, Mossor\'o-RN, Brazil\label{addr1} \and Departamento de F\'\i sica, Universidade Federal de Campina Grande, Caixa Postal 10071, 58429-900, Campina Grande-PB, Brazil\label{addr2}
}

\date{Received: date / Accepted: date}

\maketitle

\begin{abstract}
	
	We investigate the non-Gaussian effects of the Saha equation in Rindler space via Tsallis statistics. By considering a system with cylindrical geometry, we deduce the non-Gaussian Saha ionization equation for a partially ionized hydrogen plasma that expands with uniform acceleration. We demonstrate conditions for the validity of the equivalence principle within the realms of both Boltzmann-Gibbs and Tsallis statistics. In the non-Gaussian framework, our findings reveal that the effective binding energy exhibits a quadratic dependence on the frame acceleration, in contrast to the linear dependence predicted by Boltzmann-Gibbs statistics. We show that an accelerated observer shall notice a more pronounced effect on the effective binding energy for $a>0$ and a more attenuated one when $a<0$. We also ascertain that an accelerated observer will measure values of $q$ smaller than those measured in the rest frame. Besides, assuming the equivalence principle, we examine the effects of the gravitational field on the photoionization of hydrogen atoms and pair production. We show that both photoionization and pair production are more intensely suppressed in regions with a strong gravitational field in a non-Gaussian context than in the Boltzmann-Gibbs framework. Lastly, constraints on the gravitational field and the electron and positron chemical potentials are derived.
\end{abstract}

\section{Introduction} 

The Saha equation, formulated by the Indian astrophysicist Meghnad Saha in 1920 \cite{saha1920liii}, is crucial in studying the fraction of ionized atoms as a function of particle densities and temperature. This approach provides essential information about various phenomena, still not very well understood, such as the creation of neutrinos in the solar core and the estimation of light element concentrations in the early universe, among other issues in astrophysics and cosmology.  

A recent study on the Saha equation in Rindler space has been carried out by De and Chakrabarty \cite{2017Prama..88...89D}. An analysis of the photoionization of hydrogen atoms and the pair production process was performed. The authors have proved that strong gravitational fields suppress the photoionization of hydrogen atoms and also pair production at high temperatures. The so-called Rindler space refers to a reference frame undergoing a uniformly accelerated motion with respect to an inertial frame. In this spacetime, the relationship between two frames of reference, one inertial and the other non-inertial with uniform acceleration, is given by the well-known Rindler coordinates \cite{rindler2012essential}. This theoretical framework has been used in a variety of physical systems, as evidenced by the references \cite{santosfermions2019,dason2019,das2017thomas,kolekar2017gravitational,dai2016hydrogen,finster2017fermionic,mertens2016revisiting}.

Reactions like photoionization and pair production typically occur in the early universe. Here, we propose to investigate such reactions considering the local effect of gravity. For instance, pair production is most significant in high-energy environments, such as near massive objects like black holes or during energetic cosmic events. By considering primordial black holes or other astrophysical self-gravitating objects during the cosmic reionization period, we can examine how gravity affects photoionization and pair production. For the sake of simplicity, we will assume that the gravitational field is produced by such astrophysical objects. Hence, one can state that a reference frame in uniform acceleration in the absence of a gravitational field is equivalent to a frame at rest in the presence of an uniform gravitational field. In other words, according to the  equivalence principle (EP), the two situations are physically equivalent.

However, the long-range effect of gravity demands a more general approach beyond Boltzmann-Gibbs (BG) statistical mechanics. It is widely known that the BG statistics is unsuitable for describing systems with long-range interactions \cite{tsallis2023introduction}. For example, Cirto et al. investigated the \textit{Validity and Failure of the Boltzmann Weight} in the context of long- and short-range interactions ($\omega/d$) for potentials of the type $\Phi(r)\propto 1/r^{\omega}$ ($\omega\geq 0$) in a $d$-dimensional space \cite{cirto2018validity}. As a result, for short-range interactions ($\omega/d>1$), numerical analysis suggests Maxwellians for the momenta, as well as the Boltzmann weight for the energies. However, for long-range interactions ($\omega/d<1$), this pattern does not hold. In this case, numerical analysis strongly suggests $q$-Gaussians for the time-averaged momenta, as well as $q$-exponential distributions for the Boltzmann weight, thus undoubtedly falling out of the scope of BG statistical mechanics. Besides, recent publications have reported new observational evidence that suggests $q$-thermostatistics accurately describes certain aspects of astrophysical self-gravitating systems (see, for instance, \cite{sanchez2022principle,almeida2021physically}). In this paper, we will investigate the validity of the EP within the context of both BG and Tsallis statistics. Additionally, we will utilize Tsallis statistics to examine the photoionization and pair production processes in Rindler spacetime, enabling us to account for the long-range interactions involved in these processes. 

Since the original publication of Tsallis in 1988 \cite{Tsallis}, Tsallis statistics has found successful application in several fields of knowledge. Examples include astrophysics and cosmology \cite{sales2022non,zamora2022thermodynamically,teimoori2024inflation,ccimdiker2023equilibrium}, particle physics \cite{pradhan2024role,tabassam2023analysis,badshah2023systematic}, among other areas of science. This statistical framework has proved to be valuable in describing various physical systems, particularly those that exhibit anomalous behavior or long-range interactions, such as the gravitational and electromagnetic forces \cite{tsallis2023introduction}.

This work is structured as follows. In Sect. \ref{section2}, an overview of Rindler spacetime is presented. We discuss considerations on the Saha equation in Rindler space in Sect. \ref{section3}. In Sect. \ref{section4}, we introduce conditions for the validity of the EP. Next, in Sect. \ref{section5}, we determine the non-Gaussian Saha equation in Rindler space and assess their effects from the accelerated observer standpoint. In Sect. \ref{section6}, we employ the EP to analyze the photoionization of hydrogen atoms and pair production. Finally, conclusions are shown in Sect. \ref{conclusion}. Throughout the paper, we will adopt the following system of natural units: $c=k_{B}=\hbar=1$.

\section{Rindler's spacetime: brief introduction} \label{section2}

In special relativity, the relationship between the coordinates of two inertial frames $S$ and $S^{\prime}$ is given by the well-known Lorentz transformations \cite{schutz2009first}
\begin{eqnarray} \nonumber
	x^{\prime} &=& \gamma(x-vt)\;, \\ \nonumber
	y^{\prime} &=& y\;, \\ \nonumber
	z^{\prime} &=& z\;,  \\ 
	t^{\prime} &=& \gamma \left(t-vx\right)\;,
\end{eqnarray}
where $\gamma=(1-v^2)^{-1/2}$ is the Lorentz factor. Here, $S^{\prime}$ moves uniformly in the $x$-direction concerning the $S$-frame. Let's assume, now, that the frame of reference $S^{\prime}$ is moving along the $x$-direction with uniform acceleration $\alpha$ concerning $S$. In this instance, the transformations are given by the Rindler coordinates \cite{torres2006uniformly,socolovsky2014rindler}: 
\begin{eqnarray} \nonumber
	x' &=& \sqrt{x^2-t^2} - \frac{1}{\alpha}\;, \\
	t' &=& \frac{1}{2\alpha}\ln\left(\frac{x+t}{x-t}\right)\;.
\end{eqnarray}
The line element in Rindler space-time for $3+1$ dimension can be written as \cite{moller1972theory,misner1973gravitation}
\begin{equation}
	ds^2 = -\left(1+\alpha x^{\prime}\right)^2d{t^{\prime}}^2 + d{x^{\prime}}^2 + d{y^{\prime}}^2 + d{z^{\prime}}^2\;,
\end{equation}
whose metric tensor in the local reference frame $S'$ of the accelerated observer takes the form
\begin{equation}
	g_{\mu\nu} = {\rm diag}\left[ -\left(1+\alpha x^{\prime}\right)^2, 1, 1, 1\right]~.
\end{equation}

The Rindler metric is a transformation of the Minkowski metric of special relativity used to depict uniformly accelerated frames in a flat spacetime. In this spacetime, one can easily show that $g_{\mu\nu}p^{\mu}p^{\nu}$ and $g_{\mu\nu}u^{\mu}u^{\nu}$ are invariant, i.e., $g_{\mu\nu}p^{\mu}p^{\nu}=-m^2$ and $g_{\mu\nu}u^{\mu}u^{\nu}=-1$ for any frame. Here, $p^{\mu}$ and $u^{\mu}$ are the four-momentum and four-velocity, respectively. For separate timelike events, we have $ds^2=-d\tau^2$, so that the connection between the proper time $\tau$ of the particle and the coordinate time $t$ is given by 
\begin{equation}
	\frac{dt}{d\tau} = \frac{1}{\sqrt{(1+\alpha x^{\prime})^2 - v^2}}~,
\end{equation}
where $\vec{v}=d\vec{x}/dt$ is the three-velocity vector. Hence, the components of the four-velocity and four-momentum are given by
\begin{eqnarray}
	u^\mu &=& (\gamma^{*}, v^x\gamma^{*}, v^y\gamma^{*}, v^z\gamma^{*})~, \\ 
	p^\mu &=& (m\gamma^{*}, mv^x\gamma^{*}, mv^y\gamma^{*}, mv^z\gamma^{*})~,
\end{eqnarray}
in which we define $\gamma^{*}=1/\sqrt{(1+\alpha x^{\prime})^2 - v^2}$. Note that when $\alpha=0$, we recover the Lorentz factor of special relativity $\gamma^{*}=\gamma$. 

The Rindler's metric tensor $g_{\mu\nu}$ is time-independent since $x'$ is constant, meaning that the energy $-p_0$ is constant on the trajectory. To determine the energy of a single particle in Rindler spacetime, we will adopt a similar approach to the one used by Schutz \cite{schutz2009first} to ascertain the particle energy for weak gravitational fields. Using four-momentum invariance, we can write the following approximation for non-relativistic particles ($|p|\ll m$):
\begin{equation}
	(1+\alpha x')p^0 = m + \frac{p^2}{2m}~.
\end{equation}  
Using the fact that $p_0=g_{00}p^0$, the energy reads as 
\begin{equation} \label{RHalmil-nr}
	E=-p_0 = m^{\prime\prime} + \frac{p^2}{2m^{\prime}}~,
\end{equation} 
where, for the sake of simplicity, we define
\begin{equation} 
	m^{\prime} = m\left(1+\alpha x'\right)^{-1}\;\;\;{\rm and}\;\;\; m^{\prime\prime} =  m\left(1+\alpha x'\right)\;.
\end{equation} 
Note that $\alpha$ is the uniform acceleration of the frame or the uniform gravitational field in the neighborhood of the point indicated by the coordinate $x'$. This statement regards the EP. It is worth stressing that the energy in the above equation does not have an explicit time dependence. Therefore, an equilibrium statistic can be inferred without any issues.

Alternatively, this energy was also determined following the variational approach \cite{moller1972theory,misner1973gravitation}. In this context, the action integral can be written as
\begin{equation}
	S = -m \int_{a}^{b}ds \equiv \int_{t_1}^{t_2}Ldt~,
\end{equation}
where $m$ is rest mass of the particle. The Lagrangian and three-momentum in the non-inertial reference frame $S'$ are given by $L = -m/\gamma^{*}$ and $\vec{p}=\gamma^{*}m\vec{v}$, respectively. Since the Hamiltonian is $H=g_{ij}p^{i}v^{j}-L$, the single particle energy can be put as \cite{misner1973gravitation,louis2011classical,de2015particle}
\begin{equation} \label{RHalmil}
	H = m\left(1+\alpha x'\right)\left(1+\frac{p^2}{m^2}\right)^{1/2}\;.
\end{equation}  

\section{Considerations on the Saha equation in Rindler space} \label{section3} 

Some basic considerations must be made before proceeding to the derivation of the Saha equation in Rindler space. Consider the following system configuration: (i) a partially ionized hydrogen plasma (a reactive mixture of neutral hydrogen atoms, hydrogen ions, electrons and photons); (ii) cylindrical geometry: in this scenario, the plasma expands along the positive $x$-direction, which coincides with the symmetry axis of the cylinder, with uniform acceleration $\alpha$; and (iii) EP: one cannot distinguish between an accelerating frame and a uniform gravitational field.

In a partially ionized hydrogen plasma, when the photoionization rate equals the recombination rate, we may express the reaction
\begin{equation}
	H_n + \gamma \leftrightarrow H^{+} + e^{-}\;,
\end{equation}  
where $n$ denotes the hydrogen atom's energy level. In a chemical equilibrium situation, we have that $\mu_{e^-}+\mu_{H^+}=\mu_{H_n}$, since $\mu_{\gamma}=0$. In Ref. \cite{2017Prama..88...89D}, the EP is used to illustrate how the gravitational field affects the pair production and photoionization of hydrogen atoms. The Saha equation in a non-inertial frame was obtained by the authors as follows: 
\begin{equation} \label{Saha-Rindler}
	R(\alpha)=\frac{n_{H^{+}}n_e}{n_{H_n}} = G_{(e,+,n)}n_{Q_e}\exp(-\beta\varepsilon_{n}^{\prime})\;,
\end{equation}  
where $n_{Q_e} = \left(m_{e}^{\prime}/2\pi\beta\right)^{3/2}$, $G_{(e,+,n)}=g_{e}g_{+}/g_{n}$ is the degeneracy factor and $\varepsilon_{n}^{\prime} = \left(1+\alpha x'\right)\varepsilon_{n}$ is a measure of excitation energy (effective binding energy in the $S'$-frame), with $\varepsilon_{n}=m_e+m_{H^+}-m_{H_n}$ being the hydrogen ionization potential measured in the inertial $S$-frame. In addition, $\beta=1/T$, for which $T$ stands for the thermal bath temperature.  

In Ref. \cite{2017Prama..88...89D}, it was assumed that the gravitational field $\alpha$ is produced by strongly gravitating objects, such as a black hole, for instance. Thus, within a small region $x'$ away from the origin of the gravitating object, $\alpha$ can be approximated by a constant value. Comparing measurements taken by observers $S'$ and $S$, the authors have suggested that photoionization and pair production are suppressed in regions with strong gravitational fields. Notice that $\varepsilon_{n}^{\prime}$ is linearly affected by $\alpha$. In other words, it becomes more difficult to ionize a hydrogen atom in the vicinity of an astrophysical object in which $\alpha\gg 0$. These results can also be addressed from the perspective of the Unruh effect \cite{unruh1976notes}. For an accelerating observer, the lowest energy state is not a Minkowski vacuum; rather, it is perceived as a thermal bath. The temperature of this heat bath is proportional to the observer's acceleration \cite{unruh1998acceleration}, i.e.,
\begin{equation}
	T_U = \frac{\alpha}{2\pi}~.
\end{equation}
This temperature is known in the literature as the Unruh temperature. Likewise to the Unruh effect, while an observer at rest measures the binding energy $\varepsilon_{n}$,  an accelerated observer measures a shift in the energy $\varepsilon_{n}$, given by $\varepsilon_{n}^{ \prime} = \left(1+\alpha x'\right)\varepsilon_{n}$. Therefore, for $\alpha x'>0$, an accelerated observer will see that $\varepsilon_{n}^{\prime}>\varepsilon_{n}$, meaning that the spacing between hydrogen energy levels should be $\left(1+\alpha x'\right)\varepsilon_{1}/n^2$ rather than $\varepsilon_{1}/n^2$. Here, $\varepsilon_{1}$ is the hydrogen binding energy in its ground state.

\section{Conditions for the equivalence principle} \label{section4}

In this section, we will address the validity of the EP in the context of reactions in thermodynamic equilibrium. According to Schutz \cite{schutz2009first}, Einstein's EP can be stated as follows: 
\begin{quote}
	\textit{Any local physical experiment not involving gravity will have the same result if performed in a freely falling inertial frame as if it were performed in the flat spacetime of special relativity.}
\end{quote}
In the realm of the reaction $H_n + \gamma \leftrightarrow H^{+} + e^{-}$, this principle can be cast as:  
\begin{quote}
	\textit{A uniformly accelerated observer $S'$, without a gravitational field, measures the photoionization reaction at equilibrium. The results should be equivalent to measurements of the same physical process in an inertial reference frame $S$ with a uniform gravitational field.}
\end{quote}
In order to check the EP for the photoionization of hydrogen atoms within the BG framework, we shall employ $R(\alpha)$ from Eq. (\ref{Saha-Rindler}), which depicts the measurement made by a uniformly accelerated frame without a gravitational field and determine $R(\phi)$, the measure taken by an inertial observer with a uniform gravitational field. 

To ascertain $R(\phi)$, we will adopt a weak gravitational field, i.e., the gravitational potential energy of a particle is much less than its rest-mass energy ($|m\phi|\ll m$). In this regard, the ordinary Newtonian potential $\phi$ completely determines the metric, which takes the form \cite{schutz2009first}
\begin{equation}
	ds^2 = -(1+2\phi)dt^2 + (1-2\phi)(dx^2 + dy^2 + dz^2)\;,
\end{equation}
whose metric tensor in the local reference frame $S$ is given by
\begin{equation}
	g_{\mu\nu} = {\rm diag}\left[ -(1+2\phi), (1-2\phi), (1-2\phi), (1-2\phi)\right]~.
\end{equation}
Here, $\phi=-GM/x'$, where $G$ and $M$ are the Newtonian constant and the source's mass, respectively. Using four-momentum invariance, the temporal component of the four-momentum can be written as follows (for $|p|\ll m$ and $|\phi|\ll 1$):
\begin{equation}
	p^0 = m - m\phi + \frac{p^2}{2m}~.
\end{equation}  
Since $p_0=g_{00}p^0$ and keeping within the
approximations aforementioned, the energy can be put as
\begin{equation} \label{NewtonianEnergy}
	E=-p_0 = m + m\phi + \frac{p^2}{2m}~,
\end{equation} 
where the first term on the right-hand side of equation stands for the particle's rest mass, and the second and third are the gravitational potential energy and kinetic energy, respectively.  

Using the above energy, the Saha ionization equation in the presence of a stationary gravitational field can be written as
\begin{equation} \label{Saha-phi}
	R(\phi) = G_{(e,+,n)}\left(\frac{m_e}{2\pi \beta}\right)^{3/2} \exp[-\beta\varepsilon_{n}(1+\phi)]\;.
\end{equation}  
For the EP to hold, it is necessary that $R(\phi)$ and $R(\alpha)$ are equal. Therefore, since $R(\alpha)/R(\phi)=1$, we find out the following result: 
\begin{equation} \label{AlphaPhi}
	\phi = \alpha x' + \frac{3}{2}\frac{T}{\varepsilon_{n}}\ln(1+\alpha x')~.
\end{equation}
The above equation explains the effect of the observer's acceleration on the gravitational potential.  The emergence of temperature and binding energy in the second term of this equation is associated with the nature of the hydrogen photoionization reaction. These parameters naturally arise in the Saha ionization equation under study. Thus, Eq. (\ref{AlphaPhi}) depicts the condition for an inertial observer with a gravitational field and an accelerated observer without a gravitational field to agree on the measurements made for $R$. Due to these aspects, it is more suitable to designate Eq. (\ref{AlphaPhi}) as a pseudo-Newtonian potential. Besides, Eq. (\ref{AlphaPhi}) suggests that a free-falling observer will perceive a thermal effect. One possible explanation for this effect involves the Unruh effect. Since we are handling uniformly accelerated frames, it is not unexpected for a thermal effect to occur. However, considering the approximation $|\alpha x'|\ll 1$, we have that $\ln(1+\alpha x')\approx \alpha x'$. Furthermore, assuming the limit $\varepsilon_{n}\gg T$, we attain $\phi=\alpha x'$. This latter equality yields
\begin{equation} \label{equivalence}
	\vec{\alpha} = -\frac{GM}{\vec{x'}^2} = \vec{g}~.
\end{equation}
This indicates that when the conditions employed above are fulfilled, Rindler's spacetime acceleration becomes equivalent to a uniform gravitational field generated by an arbitrary astrophysical object of mass $M$, demonstrating the validity of the EP.  

\begin{figure}[!ht]
	\centering
	\includegraphics[scale=0.54]{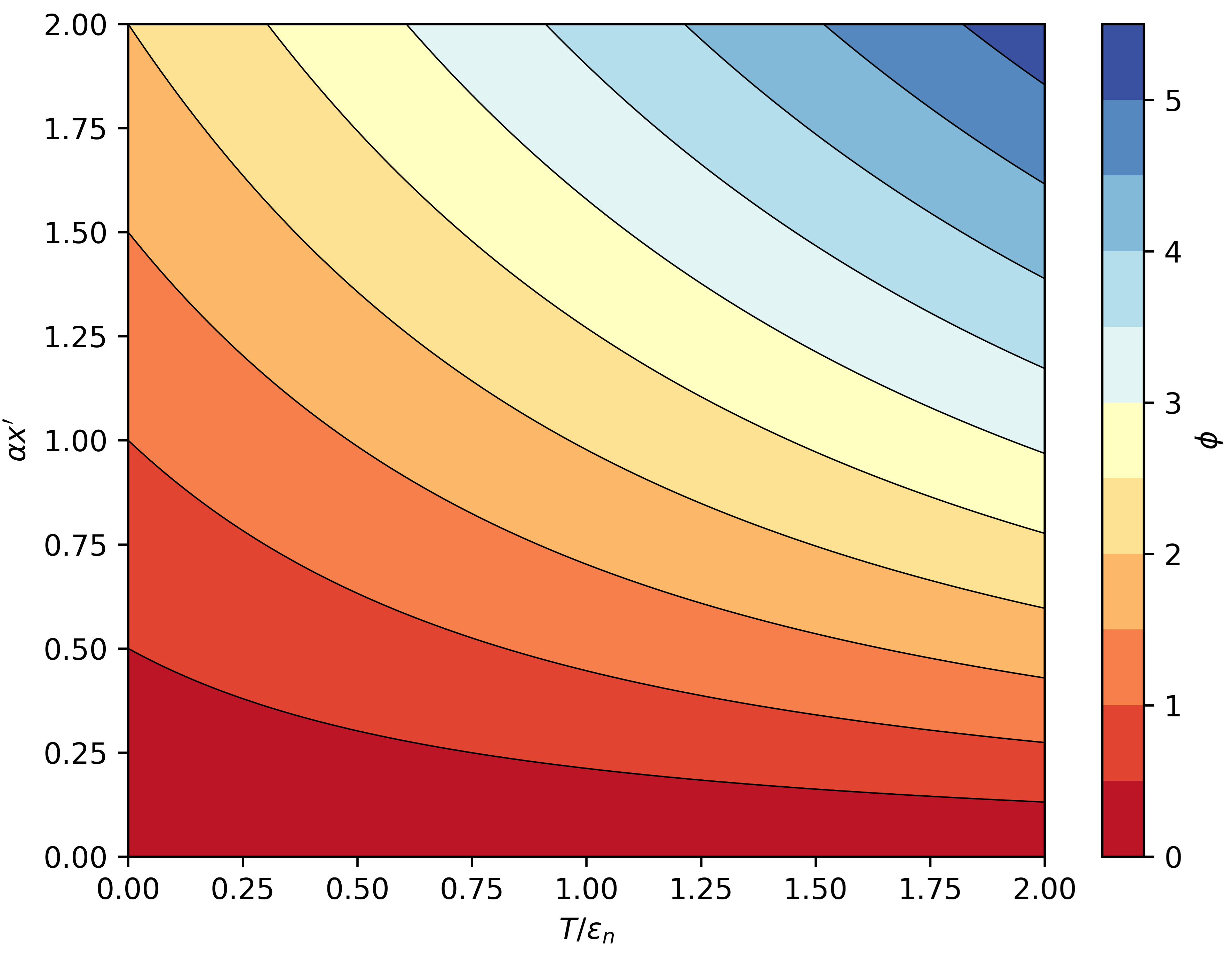}
	\caption{Contour plot for the pseudo-Newtonian potential on the $T/\varepsilon_{n} - \alpha x'$ plane. Each line illustrates a specific constant value of $\phi$}
	\label{fig1}
\end{figure}

Let's analyze the behavior of the pseudo-Newtonian potential graphically without making any assumptions in advance. Fig. \ref{fig1} shows a contour plot relating the dependent variable $\phi$ with the independent variables $\alpha x'$ and $T/\varepsilon_{n}$. It can be noticed from the graph that the potential on the $\alpha x' - T/\varepsilon_{n}$ plane raises as $\alpha x'$ increases while keeping $T/\varepsilon_{n}$ constant. Similarly, for fixed $\alpha x'$, $\phi$ increases with growing $T/\varepsilon_{n}$. Contour lines closer together indicate a rapid change in potential (e.g., when $T/\varepsilon_{n}\approx 2$), whereas those further apart represent a slow change in $\phi$ (e.g., for $T/\varepsilon_{n}\approx 0$). The region of highest potential is located in the upper right corner of the figure. Meanwhile, the lower left corner is the region with the lowest potential. Eventually, for $T/\varepsilon_{n}\approx 0$, there is a linear trend, since in this situation $\phi\approx \alpha x'$.

\section{Non-Gaussian effects of Saha's ionization in Rindler spacetime} \label{section5}

In 1988, inspired by multifractal systems, Tsallis introduced a new entropic form for the BG entropy \cite{Tsallis}. This new non-Gaussian statistical framework recovers the BG statistics as a particular case ($q\rightarrow 1$). Mathematically, the Tsallis entropy is defined as follows:
\begin{equation} \label{qentropy}
	S_{q} = \frac{1}{q-1}\left(1-\sum_{i=1}^{\Omega}p_{i}^{q}\right)~,
\end{equation}
where $p_{i}$ is the probability of the system being in the microstate $i$, $\Omega$ the total number of settings, and $q$ the parameter that measures the deviation from standard equilibrium or the degree of non-additivity of the entropy (commonly referred to as the entropic index). This parameter is also associated with the intensity of correlations of the system \cite{da2023nonextensive,de2023range} and the long-range interactions \cite{cirto2018validity}. 

In the Tsallis statistics, the logarithm and exponential functions are rewritten as power laws as follows:
\begin{equation}
	\ln_q(x)=\frac{x^{1-q}-1}{1-q}
\end{equation}
and
\begin{equation} 
	\exp_q(x) = e_{q}^{x} = [1+(1-q) x]^{\frac{1}{1-q}}\;.
\end{equation}
These modified functions return to usual functions when $q\rightarrow 1$. Many helpful properties in the Tsallis framework can be found in Ref. \cite{borges2004possible}.  

\subsection{Non-Gaussian number density in Rindler spacetime}

In the realm of Tsallis' non-Gaussian statistics, the particle number $q$-density of species $i$ might be defined as 
\begin{equation} \label{dnpeg} 
	n_{i}^{q} \equiv \frac{N_{i}^{q}}{\Delta V} = \frac{g_{i}}{(2\pi)^{3}}\int d^{3}p \mathcal{N}_{i}^{q}~,
\end{equation}
where $\mathcal{N}_{i}^{q}$ is the generalized occupation number and $g_i$ the degeneracy of species $i$. Besides, $\Delta V=A\Delta x'$ is a small volume element with $A$ being the cross-sectional area of the cylinder, and $\Delta x'$ is a small length element in the $x'$-direction at a distance $x'$ from the origin. Using a non-Gaussian fermionic distribution and neglecting the quantum effects, Eq. (\ref{dnpeg}) becomes 
\begin{equation} \label{dnpeg1}
	n_{i}^{q} = \frac{4\pi g_{i}}{(2\pi)^{3}}e_{q}^{\beta(\mu_{i}-m_{i}^{\prime\prime})}\int_{0}^{\tau} dpp^{2}e_{q}^{-\gamma_{i} p^2}\;,
\end{equation}
where $\mu$ is the chemical potential and
\begin{eqnarray} \label{a}
	\tau = \left\{
	\begin{array}{rl}
		\infty, & q>1 \\
		\left[(1-q)\gamma_{i}\right] ^{-1/2}, & q<1~.
	\end{array}
	\right.
\end{eqnarray}
Also, 
\begin{equation} 
	\gamma_{i} = \frac{\beta}{2m_{i}^{\prime}[1+(1-q)\beta(\mu_{i}-m_{i}^{\prime\prime})]}\;.
\end{equation}

The non-Gaussian number density for a non-inertial frame of reference in the non-relativistic regime is given by
\begin{equation} \label{q-density}
	n_{i}^{q} = g_{i}B_{q}n_{Q_i}\left[ e_{q}^{\beta(\mu_{i}-m_{i}^{\prime\prime})}\right]^{\frac{5-3q}{2}}~,
\end{equation}
where 
\begin{equation} 
	B_{q} = \left\{
	\begin{array}{rcl}
		\displaystyle{\frac{1}{(q-1)^{3/2}}\frac{\Gamma\left(\frac{5-3q}{2(q-1)}\right)}{\Gamma\left(\frac{1}{q-1}\right)}},& \mbox{if} & 1<q<5/3 \\ \\ 
		\displaystyle{\frac{1}{(1-q)^{3/2}}\frac{\Gamma\left(\frac{2-q}{1-q}\right)}{\Gamma\left(\frac{7-5q}{2(1-q)}\right)}}, & \mbox{if} & q<1~.
	\end{array}
	\right. 
\end{equation}
The abovementioned equations were originally derived in Ref. \cite{sales2022non} for the case $\alpha=0$. Nevertheless, in this study, we extend this framework by considering the effects of non-relativistic energy in Rindler spacetime, as shown in Eq. (\ref{RHalmil-nr}). It can be easily ascertained that these expressions revert to their form obtained in Ref. \cite{2017Prama..88...89D} when $q\rightarrow 1$.

\subsection{Non-Gaussian Saha equation in Rindler spacetime}

It is well known that Saha's equation is built under the assumption of chemical equilibrium. In the BG framework, this equation can be used to estimate concentrations or determine the free electron fraction as a function of temperature, as long as the equilibrium is maintained. Nevertheless, the presence of long-range interactions or strong statistical correlations typically drags the system into non-equilibrium states \cite{tirnakli2016standard}. As the BG statistics is inadequate for handling systems exhibiting such features, the Saha equation must account for these effects beyond the scope of BG. It can be observed in Eq. (\ref{Saha-Rindler}) that $R(\alpha)\propto \exp(-\beta \varepsilon_{n}^{\prime})$ bears a striking resemblance to the Boltzmann weight. Indeed, assuming the EP, this represents the Boltzmann weight under the local effect of gravity. Hence, based on the results of Ref. \cite{cirto2018validity}, it is apparent that $R(\alpha)\propto \exp(-\beta \varepsilon_{n}^{\prime})$ cannot accurately depict the physical phenomena in the presence of long-range effect of gravity. Therefore, it is a sensible physical assumption to generalize the Saha equation to the following $q$-exponential distribution: $R(\alpha, q)\propto \exp_{q}(-\beta \varepsilon_{q,n}^{\prime})$.

Henceforward, let us establish the Saha equation in Rindler spacetime for a partially ionized hydrogen plasma\footnote{In our approach, we will take into account the simpler situation: all parts of the fluid have the same proper acceleration, i.e., a uniform gravitational field.} in the Tsallis framework, and assess the effects of a non-Gaussian contribution. Taking $n_{i}^{q}$ for electrons ($e^{-}$), hydrogen ion ($H^{+}$), and hydrogen atoms in their $n$-th excited state ($H_n$), the non-Gaussian version of Eq. (\ref{Saha-Rindler}) becomes
\begin{equation} \label{q-Saha-Rindler}
	R(\alpha,q)=\frac{n_{H^{+}}^{q}n_e^{q}}{n_{H_n}^{q}} = G_{(e,+,n)}n_{Q_e}B_{q}\left( e_{q}^{-\beta\varepsilon_{q,n}^{\prime}}\right)^{\frac{5-3q}{2}}\;,
\end{equation}
where we set
\begin{equation} 
	\varepsilon_{q,n}^{\prime} =\frac{\varepsilon_{n}^{\prime}+(q-1)\beta m_{e}^{\prime\prime}m_{H^+}^{\prime\prime}}{1+(q-1)\beta m_{H_n}^{\prime\prime}}~, 
\end{equation}
or in terms of observer's acceleration $\alpha$ as
\begin{equation} \label{elgr}
	\varepsilon_{q,n}^{\prime} = \frac{\left(1+\alpha x'\right)\varepsilon_{n}+(q-1)\left(1+\alpha x'\right)^{2}\beta m_{e}m_{H^+}}{1+(q-1)\left(1+\alpha x'\right)\beta m_{H_n}}~.
\end{equation}
While in the BG statistics, the effective binding energy is proportional to $(1+\alpha x')$ (linear dependence), under a non-Gaussian standpoint, the effective binding energy is proportional to $(1+\alpha x')^2$ (quadratic dependency). In other words, the non-Gaussian effect on the binding energy in Rindler space is characterized by a quadratic dependence on the observer's acceleration, governed by the $q$-parameter. 

\begin{figure}[!ht]
	\centering
	\includegraphics[scale=0.54]{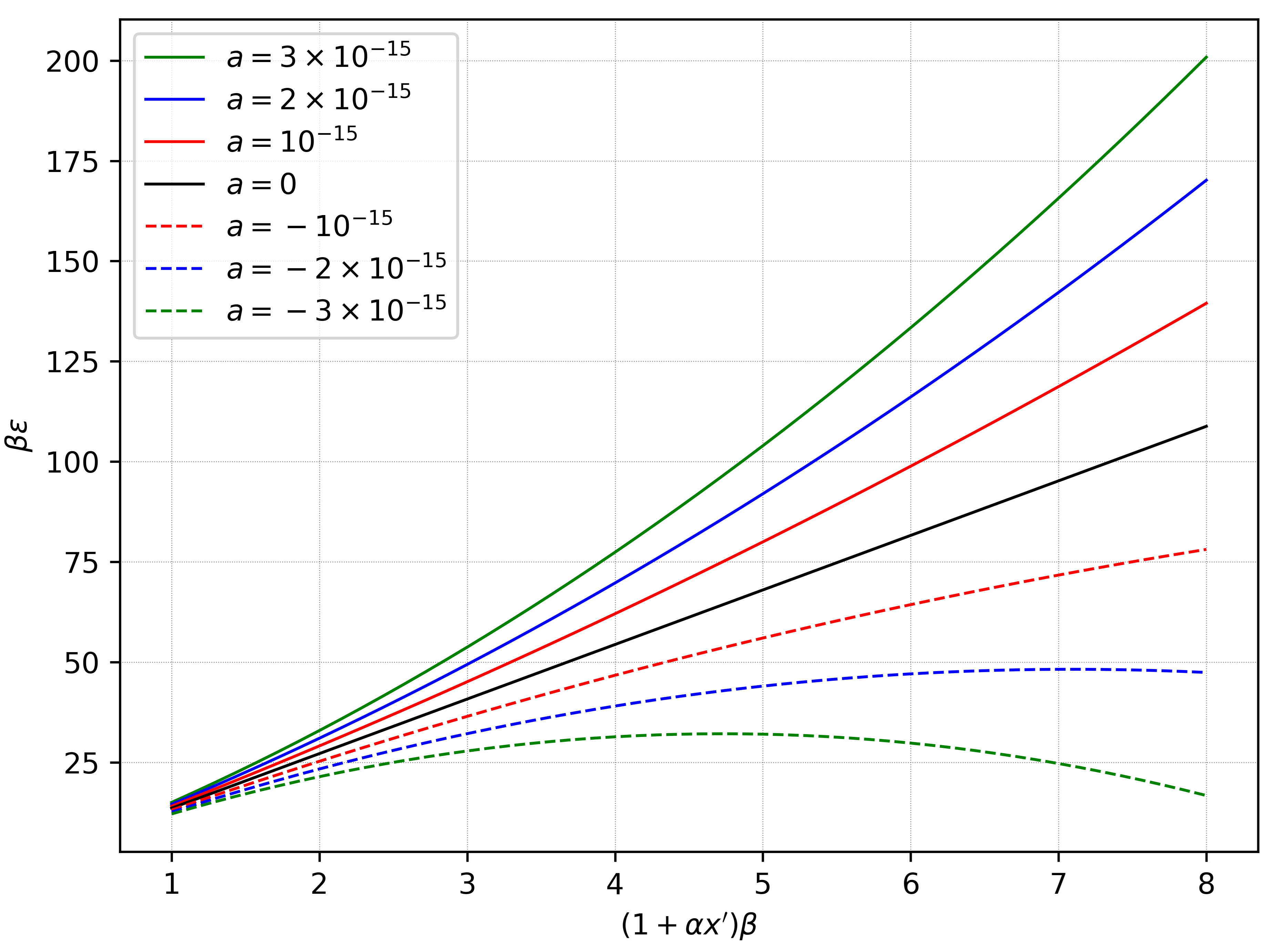}
	\caption{Effects of Tsallis statistics on the effective binding energy from the perspective of an accelerated observer. To obtain these lines, we assign the following values: $m_e=0.511~{\rm MeV}$, $m_{H^+}=938.272~{\rm MeV}$, $m_H=938.783~{\rm MeV}$ and $\varepsilon_{1}=13.598~{\rm eV}$}
	\label{fig2}
\end{figure}

Fig. \ref{fig2} portrays the effects of the Tsallis statistics on the effective binding energy for some values of the parameter $a$, defined as $a=q-1$. Notably, for $\varepsilon_n >0$, we have $\varepsilon_{q,n}^{\prime}>\varepsilon_{n}^{\prime}$ if $q>1$, and $\varepsilon_{q,n}^{\prime}<\varepsilon_{n}^{\prime}$ whenever $q<1$. Note that, at very high temperatures ($T\rightarrow \infty$), the quadratic dependence in Eq. (\ref{elgr}) disappears and then $\varepsilon_{q,n}^{\prime}\rightarrow \varepsilon_{n}^{\prime}$. Likewise, this latter outcome is also found when $q\rightarrow 1$ or equivalently $a\rightarrow 0$. The $a$-parameter plays a meaningful role in the intensity of $\varepsilon_{q,n}^{\prime}$ measured in the reference frame $S'$. Here, the effective binding energy $\varepsilon_{q,n}^{\prime}$ carries on being interpreted as a measure of excitation energy, now amplified due to the quadratic dependence of the uniform acceleration and the sensitivity to the $a$-parameter. As shown in Fig. \ref{fig2}, for very small $a$, the lines generated depict an approximately linear behavior mainly for small values of $(1+\alpha x')\beta$. Besides, as the factor $(1+\alpha x')\beta$ raises, the impact of the $a$-parameter on the effective binding energy in the non-inertial frame becomes more noticeable. Within the Tsallis framework, an accelerated observer will notice a more pronounced effect for $a>0$ and a more attenuated one if $a<0$ concerning the same observer's perception within the scope of BG statistics ($a=0$). In summary, these findings might be interpreted similarly to the Unruh effect. However, it is worth noting that the Tsallis statistics amplify the measures accomplished by an accelerated observer due to the quadratic contribution to the effective binding energy.   

\begin{figure}[!ht]
	\centering
	\includegraphics[scale=0.54]{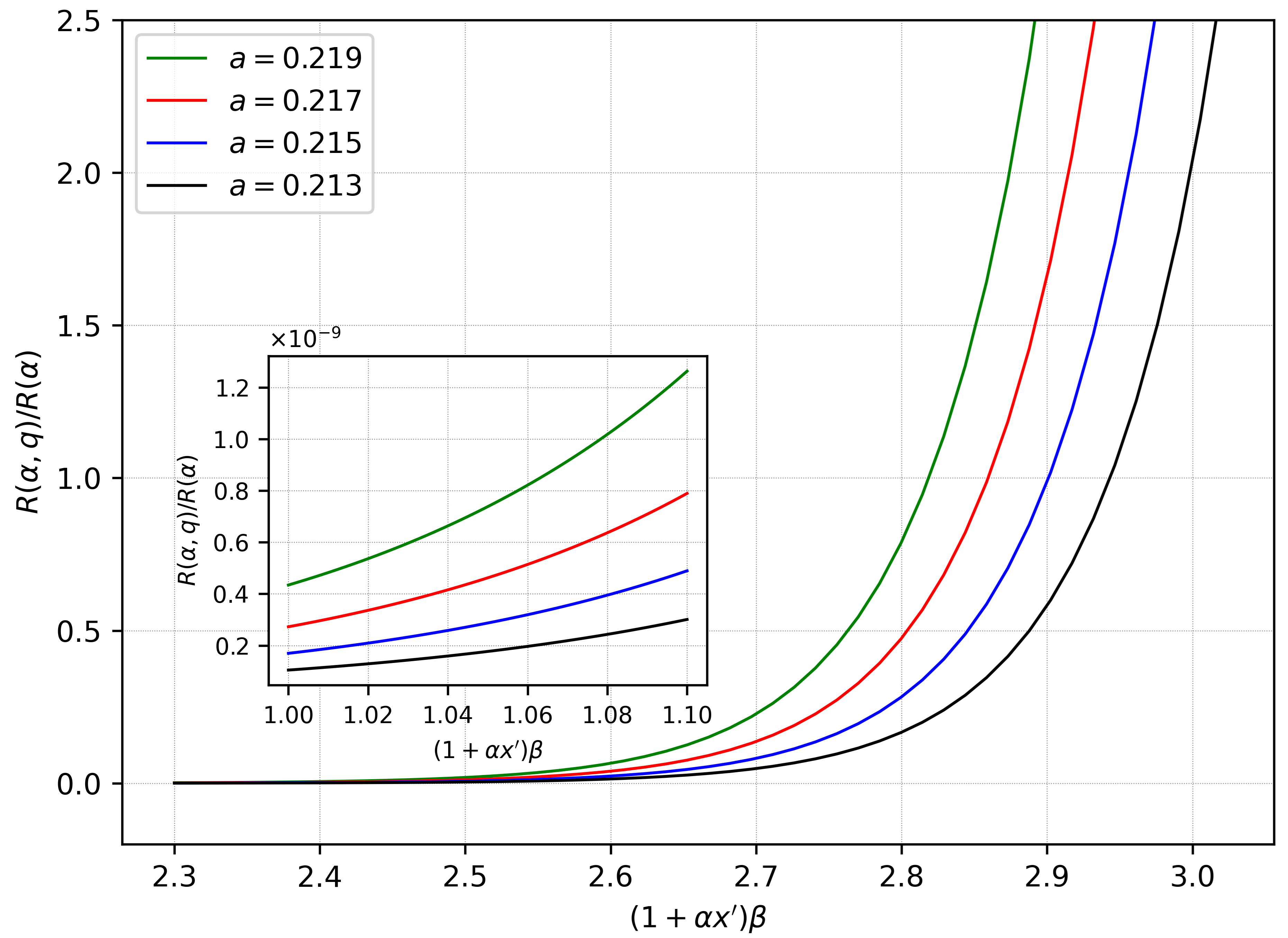}
	\caption{Perceptions of an accelerating observer regarding the measurements made for the ratio $R(\alpha,q)/R(\alpha)$ for some values of the $a$-parameter}
	\label{fig3}
\end{figure}

So far, we have looked into the non-Gaussian effects on the effective binding energy separately. Now, let us examine the magnitude of these effects in terms of concentrations measured by $R(\alpha)$ and $R(\alpha, q)$. For this purpose, consider the following ratio:
\begin{equation}
	\frac{R(\alpha, q)}{R(\alpha)} = B_q \frac{\left[ \exp_{q}(-\beta\varepsilon_{q,n}^{\prime})\right]^{\frac{5-3q}{2}}}{\exp(-\beta\varepsilon_{n}^{\prime})}~.
\end{equation}
This ratio measures the effects of the $q$-parameter on predicting concentrations for an accelerated observer. Fig. \ref{fig3} displays how this ratio behaves graphically as a function of $(1+\alpha x')\beta$ for four different values of the $a$-parameter. For the given $a$-values, we can notice that $R(\alpha,q)$ and $R(\alpha)$ are equal for certain values of $(1+\alpha x')\beta$. For instance, for $a=0.213$ and $a=0.219$, we obtain $(1+\alpha x')\beta=2.943$ and $(1+\alpha x')\beta=2.819$, respectively. This means that, even in a non-Gaussian picture, it is possible to make predictions of concentrations equal to those eventually made in the BG framework. On the other hand, for $(1+\alpha x')\beta\approx 1$, we have $R(\alpha,q)/R(\alpha)\sim 10^{-9}$, highlighting a more pronounced difference between the two approaches.  

\subsection{Equivalence principle in the Tsallis framework}

To evaluate the EP for the photoionization of hydrogen atoms in the Tsallis picture, we will use $R(\alpha, q)$ from Eq. (\ref{q-Saha-Rindler}), which represents the measurement taken by a uniformly accelerated frame with no a gravitational field. The measure made by an inertial observer with a uniform gravitational field is given by
\begin{equation} \label{q-Saha-phi}
	R(\phi,q) = G_{(e,+,n)}\left(\frac{m_e}{2\pi\beta}\right)^{3/2}B_{q}\left[e_{q}^{-\beta\varepsilon_{q,n}(\phi)}\right]^{\frac{5-3q}{2}}\;,
\end{equation}
where
\begin{equation} \label{energy-phi}
	\varepsilon_{q,n}(\phi) = \frac{\left(1+\phi\right)\varepsilon_{n}+(q-1)\left(1+\phi\right)^{2}\beta m_{e}m_{H^+}}{1+(q-1)\left(1+\phi\right)\beta m_{H_n}}~.
\end{equation}

According to EP, we must have $R(\phi,q)=R(\alpha,q)$. With this in mind, and after a little algebra, we obtain the following outcome:
\begin{equation} \label{equivalence-q}
	(\phi)_q = \frac{(\alpha x')_q + \frac{T}{\varepsilon_n}\ln_q(1+\alpha x')^{\frac{3}{5-3q}}}{1 + (1-q)\frac{T}{\varepsilon_n}\ln_q(1+\alpha x')^{\frac{3}{5-3q}}}~,
\end{equation}
for which
\begin{equation} \label{phi-q}
	(z)_q = \frac{\left(1+z\right)+(q-1)\left(1+z\right)^{2}\beta m_{e}m_{H^+}/\varepsilon_{n}                     }{1+(q-1)\left(1+z\right)\beta m_{H_n}}~.
\end{equation} 
Eq. (\ref{equivalence-q}) presents a result analogous to Eq. (\ref{AlphaPhi}) in a non-Gaussian scenario. It can be easily shown that the above equation retrieves Eq. (\ref{AlphaPhi}) in the limit $q=1$. However, under the conditions imposed in Sect. \ref{section4}, i.e. $|\alpha x'|\ll 1$\footnote{In this case, we have that $\ln_q(1+\alpha x')^{3/(5-3q)}\approx 3\alpha x'/(5-3q)$.} and $\varepsilon_{n}\gg T$, we achieve $(\phi)_q=(\alpha x')_q$, which implies $\varepsilon_{q,n}(\phi)=\varepsilon_{q,n}^{\prime}(\alpha)$. This latter equality suggests that the effective binding energy remains consistent between an inertial frame experiencing a gravitational field and a uniformly accelerated observer in the absence of a gravitational field. Hence, the solely possible solution for $\varepsilon_{q,n}(\phi)=\varepsilon_{q,n}^{\prime}(\alpha)$ is $\phi=\alpha x'$, demonstrating the validity of the EP within the scope of Tsallis statistics. It is worth assessing the temperature extremes in light of Eq. (\ref{equivalence-q}) and comparing them with Eq. (\ref{AlphaPhi}). Note that the equivalence principle remains valid for both approaches at $T=0$, i.e., $\phi_{\rm BG}=\phi_{\rm Tsallis}=\alpha x'$. On the other hand, as $T\rightarrow \infty$, we observe $\phi_{\rm BG}\rightarrow \infty$ and $\phi_{\rm Tsallis}=q/(1-q)$. Notably, the limits are compatible when $q\rightarrow 1$.

A link between $q$ and $\alpha$ can be obtained from the effective binding energy, Eq. (\ref{elgr}). We have already seen that an observer accelerated under the BG statistics will notice the following shift in the hydrogen binding energy: $\varepsilon_{n}^{\prime}=(1+\alpha x')\varepsilon_1/n^2$. Suppose we want to figure out the values of the $q$-parameter that allow us to access other hydrogen energy levels in the accelerated reference frame, assuming that $\varepsilon_{q,n}^{\prime}=(1+\alpha x')\varepsilon_1/n^2$. Accordingly, we might to establish the following relationship: 
\begin{equation} \label{qprime}
	(q-1)^{\prime} = \frac{(q-1)}{(1+\alpha x')}~.
\end{equation}
From the perspective of an accelerated observer, these are the values of the factor $q$ required to access other hydrogen energy levels. The measurement made in the rest frame ($q-1$) was illustrated in Ref. \cite{sales2022non}.  It can be remarked in the equation above that $(q-1)^{\prime}$ is inversely proportional to $(1+\alpha x')$, i.e., as $\alpha$ increases $(q-1) ^{\prime}$ decreases. Then, for $\alpha x'>0$, an accelerated observer will measure values of $q$ smaller than those measured in the rest frame. Also, there is a singularity in Eq. (\ref{qprime}) when $\alpha x'=-1$. This leads to a constraint on the distance $x'$. Hence, since the EP claims that $\phi=\alpha x'$, we find $x'\neq GM$.

Lastly, a meaningful remark is that we can obtain a constraint for the acceleration $\alpha$ according to the following property:
\begin{equation} \label{property}
	\frac{\exp_{q}(x)}{\exp_{q}(y)} = \exp_{q}\left[\frac{x-y}{1+(1-q)y}\right]~~~~\left(y\neq \frac{1}{q-1}\right)~.
\end{equation} 
Thus, Eq. (\ref{q-Saha-Rindler}) is constrained as
\begin{equation} \label{constraint-1}
	\alpha \neq \frac{1}{m_{H_n}x'}\left(\mu_{H_n}-m_{H_n}-\frac{T}{q-1}\right)\;.
\end{equation}
This implies that the non-Gaussian effects on photoionization in Rindler space only hold if we take into account the constraint given by Eq. (\ref{constraint-1}).

\section{Applications} \label{section6}   

\subsection{Photoionization}

In this section, we shall examine the photoionization of hydrogen atoms taking into account the EP. For this end, consider the ratio with $\alpha>0$ and $\alpha=0$:
\begin{equation}
	I(\alpha,q) \equiv \frac{R(\alpha,q)}{R(0,q)} = \left(1+\alpha x'\right)^{-3/2}\left(e_{q}^{-\beta\Phi_q}\right)^{\frac{5-3q}{2}}\;,
\end{equation} 
where $I$ measures how the photoionization of hydrogen atoms is affected by the gravitational field. Besides, we define
\begin{equation} \label{phiq}
	\Phi_q = \frac{\varepsilon_{q,n}^{\prime}-\varepsilon_{q,n}}{1+(q-1)\beta\varepsilon_{q,n}}\;.
\end{equation}
We have demonstrated in Eq. (\ref{elgr}) that $\varepsilon_{q,n}^{\prime}\propto (1+\alpha x')^2$. Therefore, the term $\Phi_q$ also has a quadratic dependency of the gravitational field. This suggests that photoionization of hydrogen atoms in Rindler space is suppressed more strongly in regions with a strong gravitational field in Tsallis statistics than in the classic one. In the limit $q\rightarrow 1$, the linear dependency is recovered, i.e., 
\begin{equation}
	I(\alpha,1) = \left(1+\alpha x'\right)^{-3/2} \exp(-\beta \alpha x'\varepsilon_n)\;.
\end{equation}
This result was also found in the work of De and Chakrabarty \cite{2017Prama..88...89D}.

\subsection{Electron-positron pair production}

In order to investigate the electron-positron pair production, consider the following reaction:
\begin{equation}
	\gamma + \gamma \leftrightarrow e^- + e^+\;.
\end{equation} 
Here, the chemical equilibrium reads as $\mu_{e^-}+\mu_{e^+}=0$, since $\mu_{\gamma}=0$. In this situation, the particle number $q$-density is written as 
\begin{equation} 
	n_{e^\mp}^{q} = g_{e}B_{q}n_{Q_e}\left\lbrace \exp_{q}\left[\beta(\mu_{e^\mp}-m_{e^\mp}^{\prime\prime})\right]\right\rbrace ^{\frac{5-3q}{2}}\;.
\end{equation}  
Hence, the product of electron-positron concentration is given by
\begin{equation}
	C(\alpha,q) \equiv n_{e^-}^{q}n_{e^+}^{q} = g_{e}^{2}B_{q}^{2}n_{Q_e}^{2}[\exp_{q}(\xi_q)]^{\frac{5-3q}{2}}\;, 
\end{equation}
where we set
\begin{equation}
	\xi_q(\alpha) = (1-q)\beta^{2}\left[ \left(1+\alpha x'\right)^{2}m_{e}^{2}-\mu_{e}^{2}\right] -2\beta \left(1+\alpha x'\right)m_{e}\;,
\end{equation}
and we employ the chemical equilibrium condition.

The ratio between the product of electron-positron concentrations with $\alpha>0$ and $\alpha=0$ can be written as follows:
\begin{equation} \label{cgq}
	\frac{C(\alpha,q)}{C(0,q)} = \left(1+\alpha x'\right)^{-3} \left(e_{q}^{\lambda_q}\right)^{\frac{5-3q}{2}}\;,
\end{equation}
where we define
\begin{equation}
	\lambda_q = \frac{\xi_{q}(\alpha)-\xi_{q}(0)}{1+(1-q)\xi_{q}(0)}\;.
\end{equation}
It is worth noting that Eq. (\ref{cgq}) also portrays a quadratic dependence of the gravitational field, as seen in the term $\xi_q(\alpha)$. Thus, the effect of $\alpha$ on pair production is analogous to that on the photoionization process. In other words, pair production is more intensely suppressed in regions with a strong gravitational field in a non-Gaussian context than in the BG framework. 

As an immediate mathematical consequence, using the property (\ref{property}), Eq. (\ref{cgq}) is constrained by imposing the following condition: 
\begin{equation}
	\mu_{e^+} \neq \left[m_{e} - \frac{T}{1-q}\left(\frac{T}{q-1} + 2m_{e}\right)\right]^{1/2}\;. 
\end{equation} 
This is a constraint to the electron and positron chemical potentials, since $\mu_{e^+}=-\mu_{e^-}$. When $q\rightarrow 1$, Eq. (\ref{cgq}) takes the form 
\begin{equation}
	\frac{C(\alpha,1)}{C(0,1)} = \left(1+\alpha x'\right)^{-3} \exp(-2\beta \alpha x' m_e)\;,
\end{equation}
as can one verifies in Ref. \cite{2017Prama..88...89D}. 

\section{Conclusions} \label{conclusion} 

In this study, we have investigated the non-Gaussian effects of the Saha ionization equation in Rindler spacetime via Tsallis statistics. Firstly, we have determined conditions for the validity of the EP within the realms of both BG and Tsallis statistics. By applying the approximations $|\alpha x'|\ll 1$ and $\varepsilon_{n}\gg T$ in the pseudo-Newtonian potential from both approaches, we have shown that Rindler's spacetime acceleration is equivalent to a uniform gravitational field generated by an arbitrary astrophysical object, affirming the validity of the EP. In addition, we have examined the temperature extremes, noting that the EP holds for both approaches at $T=0$. On the other hand, while the potential diverges in the BG framework for $T\rightarrow \infty$, in the Tsallis picture, it converges to $\phi_{\text{Tsallis}}=q/(1-q)$. 

Additionally, we have ascertained that the effective binding energy exhibits a quadratic dependence on the frame acceleration, in contrast to the linear dependence predicted by BG statistics. Therefore, an accelerated observer shall notice a more pronounced effect on the effective binding energy for $a>0$ and a more attenuated one when $a<0$. In other words, the Tsallis statistics amplify the measures taken by an accelerated observer due to the quadratic contribution to the effective binding energy. We have also evaluated the effects of the $a$-parameter on the concentrations measured by $R$. It was shown that, even in a non-Gaussian context, there are values of the $a$-parameter that make predictions equal to ones made in the BG framework. Besides, we have explored the possibility of accessing other energy levels of the hydrogen atom and shown that an accelerated observer will measure values of $q$ smaller than those measured in the rest frame. As applications, our analysis has provided insight into the photoionization of hydrogen atoms and electron-positron pair production. We have adopted the EP to assess the impacts of the gravitational field on these physical processes in a non-Gaussian context. Our findings demonstrate that both photoionization and pair production are more intensely suppressed in the presence of a gravitational field in a non-Gaussian context than in the Boltzmann-Gibbs framework. Furthermore, we have derived constraints on the gravitational field and the electron and positron chemical potentials. 

Tsallis statistics have been successfully used to describe non-equilibrium states or long-range interactions and correlations. For instance, observations of the $q$-triplets by Voyager 1 in the solar wind indicate the manifestation of non-Gaussian effects due to non-equilibrium states \cite{burlaga2005triangle}. This solar wind is a driven non-linear non-equilibrium system. Moreover, the long-range interactions present in several physical phenomena suggest that Tsallis' statistical mechanics yields results beyond the scope of classical statistics (see \cite{plastino2022tsallis,bagchi2016sensitivity,christodoulidi2016dynamics}). In astrophysical self-gravitating systems like Polytropes stars, which are systems in thermal meta-equilibrium, the results are compatible with the principle of maximum Tsallis entropy, as discussed in Ref. \cite{almeida2021physically}. In conclusion, non-Gaussian effects should be taken into account when analyzing the behavior of a partially ionized hydrogen plasma in the presence of long-range interactions caused by the gravitational field.

\begin{acknowledgements}
	The authors sincerely thank C. A. Wuensche for his valuable comments and suggestions. LLS is thankful to CAPES/FAPESQ for financial support. FCC was supported by CNPq/FAPERN/PRONEM.
\end{acknowledgements}


\bibliographystyle{spphys}       
\bibliography{mybibfile}

\end{document}